\documentclass[reprint,pra,twocolumn,aps,showpacs,superscriptaddress,floatfix]{revtex4-1}
\usepackage[latin1]{inputenc}
\usepackage{bm}
\usepackage[usenames]{color}
\usepackage{multirow}
\usepackage{amssymb}
\usepackage{amsbsy}
\usepackage{amsmath}
\usepackage{stmaryrd}
\usepackage{graphicx}
\usepackage{epsfig}
\usepackage{placeins}
\makeatletter

\usepackage[plainpages=false,pdfpagelabels,colorlinks=true,linkcolor=red,urlcolor=blue,citecolor=blue,pdftitle={Title},pdfauthor={},pdfdisplaydoctitle=true,pdfduplex=DuplexFlipLongEdge]{hyperref}

\definecolor{darkgreen}{rgb}{0,0.60,.2}

\begin{document}

\title{Unveiling hidden structure of many-body wavefunctions of integrable systems via sudden expansion experiments}

\author{Zhongtao Mei}
\affiliation{Department of Physics, University of Cincinnati, Cincinnati, Ohio 45221-0011, USA}
\author{L. Vidmar} 
\author{F. Heidrich-Meisner}
\affiliation{Department of Physics and Arnold Sommerfeld Center for Theoretical Physics,
Ludwig-Maximilians-Universit\"at M\"unchen, 80333 M\"unchen, Germany}
\author{C. J. Bolech}
\affiliation{Department of Physics, University of Cincinnati, Cincinnati, Ohio 45221-0011, USA}

\date{\today}

\begin{abstract}
In the theory of Bethe-ansatz integrable quantum systems, rapidities play an important role as they are used to specify many-body states, apart from phases.  
The physical interpretation of rapidities going back to Sutherland is that they are the asymptotic momenta after letting a quantum gas expand into a larger volume making it dilute and noninteracting.
We exploit this picture to make a direct connection to quantities that are accessible in sudden-expansion experiments with ultracold quantum gases. 
By a direct comparison of Bethe-ansatz and time-dependent density matrix renormalization group results, we demonstrate that the expansion velocity of a one-dimensional Fermi-Hubbard model can be predicted from knowing the distribution of occupied rapidities defined by the initial state.
Curiously, an approximate Bethe-ansatz solution works well also for the Bose-Hubbard model. 
\end{abstract}
\maketitle

{\it Introduction.}
Some often emphasized aspects of experiments with ultracold quantum gases as compared to condensed matter systems are the high degree of tunability of dimensionality and interaction strengths and the possibility to obtain clean realizations of standard many-body Hamiltonians such as the Fermi- and Bose-Hubbard model~\cite{greiner02,joerdens08,schneider08,hart14}, albeit usually with inhomogeneous density profiles~\cite{bloch08}.
Moreover, ultracold quantum gases are complementary to condensed-matter systems with respect to the typically accessible observables. 
Primarily, these are in-situ density profiles or density profiles after time-of-flight experiments (possibly combined with other perturbations of the system or quenches of parameters). 
A time-of-flight measurement, achieved by turning off all potentials, gives access to either the in-situ momentum distribution function or, if the system was initially in a lattice and the removal of the lattice occurred sufficiently adiabatically, the quasi-momentum distribution. 
This relies on various assumptions~\cite{gerbier08,vidmar15}, including that the atoms do not experience interactions during the time-of-flight measurement.

This latter assumption may either not always be valid or one may actually deliberately be interested in elucidating the very
effect of interactions, which thus has a very different purpose from a time-of-flight experiment.
Here, we address two questions, namely, first, whether the density profiles during the expansion contain any information about the initial state and, second, what is the form of the asymptotic momentum distribution function.
%To clearly distinguish from the time-of-flight measurement where atoms propagate in free space, we are interested in dynamics in the lattice and with interactions, often referred to as a sudden expansion.
Theoretically, in higher dimensions, the analysis is mostly based on time-dependent mean-field schemes~\cite{castin96} (see~\cite{zaletel15,schlunzen16} for more recent methodological developments, though). 
In one dimension, for situations that either permit a scaling solution (see, e.g.,~\cite{delcampo06,gritsev10}) or are described by integrable Hamiltonians such as the Lieb-Liniger model of interacting bosons in the continuum, these questions can in principle be answered exactly~\cite{ohberg02,delcampo08,buljan08,jukic08,jukic09,minguzzi05,iyer12}.
We are particularly interested in an integrable lattice model with repulsive onsite interactions, the Fermi-Hubbard model (FHM)~\cite{essler-book}
\begin{equation}
H= -J \sum_{i=1}^{L-1} \sum_{\sigma \in \{ \downarrow, \uparrow\}} (c_{i+1\sigma}^{\dagger}c_{i\sigma} + \mathrm{h.c.}) + U \sum_{i=1}^L n_{i\uparrow} n_{i\downarrow}\,,
\end{equation}
where  $L$ is the number of lattice sites, $c_{i\sigma}$ is a fermion annihilation operator, and $n_{i\sigma} = c_{i\sigma}^\dagger c_{i\sigma}$.
In the context of bosons, the simplest, completely understood example are hard-core bosons (HCBs, the lattice version of the Tonks-Girardeau gas~\cite{kinoshita04,paredes04}), which can be obtained from the Bose-Hubbard model (BHM)
\begin{equation}
H=-J \sum_{i=1}^{L-1} (a^{\dagger}_{i+1} a_i + \mathrm{h.c.}) + \frac{U}{2} \sum_{i=1}^L n_i (n_i-1)
\end{equation}
by sending $U/J\to \infty$ or requiring $(a^{\dagger}_{i})^2=0$.
Here, $a_{i}$ is a boson annihilation operator and $n_{i} = a_i^\dagger a_i$.
For HCBs, the physical quasimomentum distribution function (quasi-MDF) $n_k$ (defined as the Fourier transform of one-particle correlations $\langle a_i^{\dagger} a_j \rangle$) after a long expansion time becomes identical to the conserved 
set $n_k^f$ of the spinless, noninteracting  fermions that HCBs can be mapped to~\cite{cazalilla11}. 
This process goes under the name of dynamical fermionization \cite{rigol05,minguzzi05} and generalizations apply to the (integrable) repulsive Lieb-Liniger gas \cite{jukic09,iyer12} and even to the (nonintegrable) BHM with $U/J<\infty$ \cite{vidmar13}.
Density profiles  $\langle n_i(t) \rangle$  undergo a ballistic expansion for HCBs in a 1D lattice, which was observed in experiments~\cite{ronzheimer13}.
The ballistic expansion manifests itself in a linear increase $R(t)=v_{\rm r} t$ of the radius defined as
\begin{equation}
R^2(t) = \frac{1}{N} \sum_i (i-i_0)^2 \langle n_i(t) \rangle\, ,
\end{equation}
where $N$ denotes the number of particles and $i_0 = (L+1)/2$.
The expansion velocity $v_{\rm r}$ of HCBs is related to the conserved $n_k^f$ and thus also to the asymptotic form of the physical $n_k(t=\infty)=n_k^f$ via \cite{langer12,vidmar13,ronzheimer13}
\begin{equation} \label{def_v}
v_{\rm r}^2 =\frac{1}{N} \sum_k (v_k)^2 n_k^f\,
\end{equation}
where $v_k = d \epsilon_k/dk $ are the group velocities of noninteracting particles with a tight-binding dispersion $\epsilon_k=-2J \cos( k)$ (the lattice spacing is set to unity).

In this work, we aim at the generalization of  these observations to Bethe-ansatz (BA) integrable lattice models with repulsive interactions that do not map onto noninteracting particles.
Following Sutherland~\cite{sutherland_98}, for such systems, quasimomenta are replaced by so-called rapidities, which have the  
interpretation that they become physical momenta in the asymptotic regime of an expansion.
This happens
once particles have spatially 
rearranged themselves
according to  increasing momenta and thus stop crossing each other as they
continue to expand.
If we then define a distribution $n_\kappa$ of rapidities $\kappa$ defined by the initial condition, then our hypothesis is that
the asymptotic physical momentum distribution function $n_k(t\to\infty)$ becomes equal to the conserved $n_{\kappa}$ (assuming, for simplicity, real $\kappa)$ 
\begin{equation}
n_k(t\to \infty) = n_{\kappa}\,.
\label{eq:nk}
\end{equation} 
As a consequence, since in the asymptotic regime the expansion is expected to be ballistic because diluteness suppresses any scattering,
we expect that the asymptotic expansion velocity can be written as
\begin{equation}
v_{\rm r}^2(t=\infty) =\frac{1}{N} \sum_\kappa (v_\kappa)^2 n_{\kappa}.
\label{eq:vr}
\end{equation}

The main result of our work is that Eq.~\eqref{eq:vr} indeed holds for the FHM with repulsive interactions and initial densities $n_0 = N/L_0$ smaller or equal than one, expanding from the correlated ground state within a box of size $L_0$
(the regime of initial filling larger than one was studied numerically in Refs.~\cite{hm09,kajala11}).
Our results are based on a comparison of a BA calculation of $n_{\kappa}$ with numerical results for the density profiles obtained from time-dependent density matrix renormalization group (tDMRG) calculations~\cite{vidal04,daley04,white04}.
This implies that a measurement of density profiles, accessible in quantum-gas experiments, gives access to the very abstract concept of rapidities of an integrable quantum  model, which are very important for actually carrying out calculations, but which are usually hidden. 
Of course, $v_{\rm r}$ is just a single number and contains only partial information about the full rapidity distribution.

A possible obstacle could be that the times needed to reach the asymptotic regime are not accessible to either experiments or tDMRG. 
This is true for the quasi-MDF \cite{hm08,bolech12}, which, using tDMRG, we are only able to obtain for $N=2,4$ particles in the long-time limit. The expansion, however, turns out to be ballistic to a good approximation (i.e., $R\propto t$ \cite{supmat}) for the FHM  under the aforementioned conditions \cite{langer12}. 
Such a behavior implies that $v_{\rm r}$ becomes time independent very rapidly  and it can thus be extracted already from short-time dynamics, long before $n_k$ has converged to its asymptotic regime.
Thus, experiments do not need to reach the asymptotic regime.

An example, in which $n_k$ nevertheless becomes stationary very fast, is the spin-imbalanced FHM with attractive interactions \cite{lu12,bolech12}, where a quantum-distillation process~\cite{hm09,muth12,xia15} ensures a fast separation of pairs and excess fermions \cite{bolech12}. In that case, the generalization of Eq.~\eqref{eq:nk} to both real and complex rapidities 
(the latter present because of the bound states in the ground state of the attractive FHM)
seems to hold, based on a comparison of tDMRG and BA calculations~\cite{bolech12}.
Interestingly, we will show here that even for the nonintegrable BHM,  one can exploit a BA approach along the lines of~\cite{krauth91}
to define $n_{\kappa}$, which via Eq.~\eqref{eq:vr} leads to a good agreement with  tDMRG data from \cite{vidmar13}.

\begin{figure}[t]
\centering
\includegraphics[width=0.99\columnwidth]{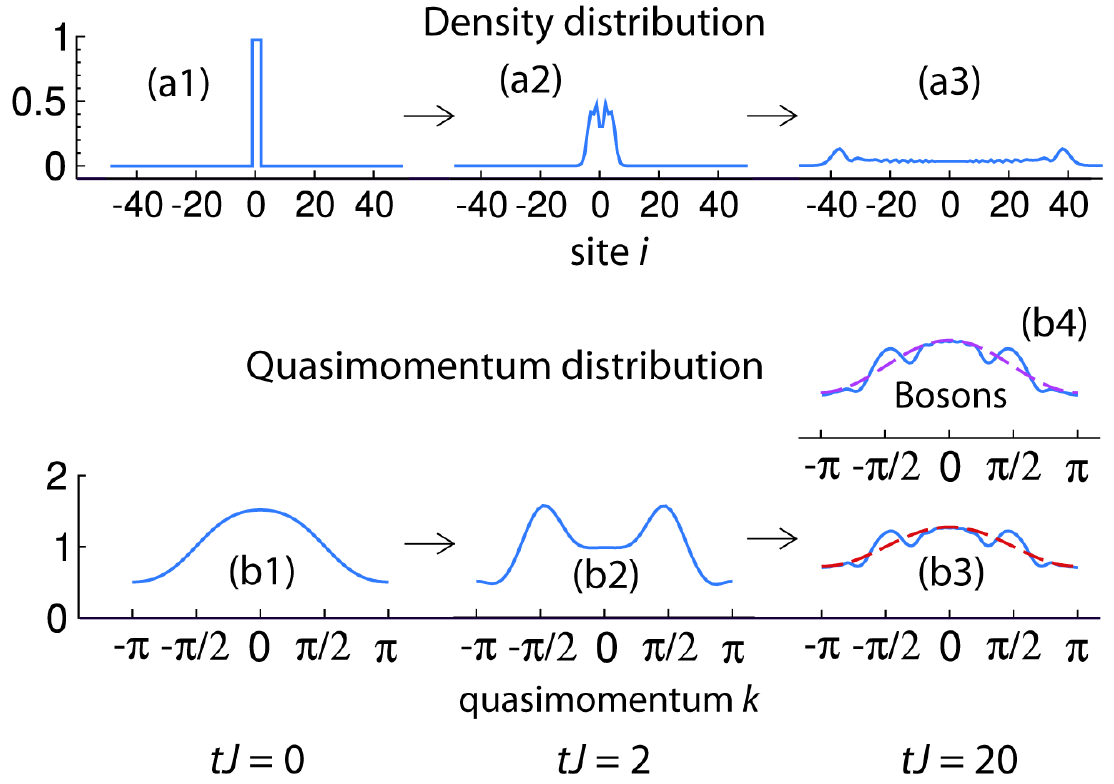}
\caption{(Color online)
{\it Sudden expansion in the FHM}. 
(a1)-(a3) Density distribution $\langle n_i(t)\rangle$ and (b1)-(b3) renormalized quasi-MDF $n_k(t)$ at (a1), (b1) $t=0$, (a2), (b2) $tJ=2$, and (a3), (b3) $tJ=20$ 
[$U=8J$, $N=4$, $L_0=N$, we set $\hbar=1$].
(b4)  Results for the BHM and the same model parameters.
Solid lines are tDMRG results, dashed lines in (b3) and (b4) show the corresponding Fermi-Dirac function, Eq.~(\ref{highT}).
In the figures, all quantities are expressed in dimensionless units.
}\label{fig:sketch}
\end{figure} 

{\it Asymptotic form of the quasi-MDF at half filling.}
We begin by describing the overall time evolution of densities and the quasi-MDF, and for the latter, we propose a simple expression for its asymptotic form for the example of initial
states with half filling.
Figure~\ref{fig:sketch} shows typical results for the FHM at $U/J=8$ obtained with tDMRG for the density profiles $\langle n_i(t)\rangle$ and the quasi-MDF $n_k(t)$.
We calculated the observables in the initial state [Figs.~\ref{fig:sketch}(a1) and~\ref{fig:sketch}(b1)], in the transient regime of the expansion [Figs.~\ref{fig:sketch}(a2) and~\ref{fig:sketch}(b2)] and in the asymptotic regime, where $n_k$  approaches its stationary form [Figs.~\ref{fig:sketch}(a3) and~\ref{fig:sketch}(b3)].
The transient regime is characterized by peaks in the quasi-MDF at $k=\pm \pi/2$~\cite{hm08}. Similar transient phenomena for bosons were studied in \cite{rigol04,rodriguez06,vidmar13,vidmar15}. 

In the long-time limit, the quasi-MDF of a gas that has expanded from a Mott insulator approaches a particle-hole symmetric form, both for fermions and bosons~\cite{vidmar13}.
For the BHM, this can be viewed as a generalized dynamical fermionization~\cite{vidmar13}, similar to  integrable bosonic models~\cite{rigol05,minguzzi05,iyer12,campbell15}.
The density profile at the longest times reached in the simulations is practically flat, except for the propagating wavefronts. Therefore, the gas can be well 
approximated by assuming both diluteness and a homogeneous density. 
We  find that the final quasi-MDF approaches a simple Fermi-Dirac distribution 
\begin{equation} \label{def_fd}
f_k = \frac{1}{e^{ \beta(\epsilon_k-\mu)}+1},
\end{equation}
where temperature $T=1/\beta$ and chemical potential $\mu$ are determined to match the  energy $E$, which is conserved during the expansion, and the particle number $N$ of the strongly correlated system~\cite{langer12}.
This effective noninteracting gas, containing the same number of particles, can be viewed as having originated from the same box.

For large $U/J$, the total energy corresponds to relatively high temperatures of the free fermions and, in addition, $\mu=0$ at half filling.
This simplifies the expression of Eq.~(\ref{def_fd}) since the only parameter that determines the quasi-MDF of the free particles $f_k$ is  the energy density $\varepsilon= |E|/N$.
Expressing the quasi-MDF up to ${\cal O}(\varepsilon^3)$, we get
\begin{equation} \label{highT}
f_k =\frac{N}{L} \left(1+ \varepsilon \cos k  - \frac{1}{3} \varepsilon^3 \left( \cos^2 k - \frac{3}{4} \right) \cos k  \right).
\end{equation}
Note that the first two terms are similar to the results of \cite{schlunzen16}.
We use Eq.~(\ref{highT}) to compare to the tDMRG result at $N=4$ for the FHM ($\varepsilon=-0.279J$) and the BHM ($\varepsilon=-0.364J$) in Figs.~\ref{fig:sketch}(b3) and~\ref{fig:sketch}(b4), respectively.
The numerical data away from $k=\pm \pi/2$ agree very well with the free-fermion reference system.
The deviation between the two curves can be quantified by looking at the difference in average velocities,
$\Delta v_{\rm av}(t) = v_{\rm av}(t) - v_{\rm av}^{(0)}$, where $v_{\rm av}^{(0)} = \sqrt{2}J$ is the equilibrium average velocity of the free-fermion system
and $v_{\rm av}(t)^2 = (1/N) \sum_k (v_k)^2 n_k(t)$.
We find that $\Delta v_{\rm av}(t)$ goes to zero at asymptotic times with a power-law dependence.

While the interpretation of the asymptotic properties in terms of a thermal state is very intuitive, it is not based on rigorous arguments, unlike the BA approach that we will detail next.
Nevertheless, the measurement of the MDF in an experiment would be very interesting and would obtain the entire rapidity distribution. 
Using few particles (as was studied in a recent  quantum-walk experiment of bosons \cite{greiner2015})  leads to a faster convergence towards the stationary form.

{\it Bethe-ansatz based approach.} We turn now to the problem of determining the asymptotic expansion velocity from Eq.~(\ref{eq:vr}). 
Generally speaking, we are attempting to predict the asymptotic form of observables from integrability in a specific quantum quench problem,
which is a question that is currently being studied for many other quenches, driving methodological advances in the framework of the Bethe ansatz  (see, e.g., \cite{iyer12,caux13,wouters14}).
We first need to determine the distribution $n_\kappa$, which is conserved for $t>0$.  
The main technical complication is that the wavefunction needs to be expanded in the postquench eigenstates (after quenching the trapping potential to zero) of the integrable homogeneous FHM in an infinitely large lattice. 
This problem is notoriously difficult, and we therefore compute the discrete set of prequench rapidities with respect to the initial box (i.e., for $t<0$, what we denote as ''rawBA'') and then use single-particle projection techniques to approximate $n_{\kappa}$ for $t>0$ (''projBA'').
The calculation of the discrete set of rapidities for particles in a box  is well defined and can be done exactly, albeit numerically (see \cite{bolech13,supmat} for details). 
Next, as the trapping potential is suddenly turned off, the initial distribution $n_\kappa(t=0^-)$ is projected into a modified one $n_\kappa(t=0^+)$ that is consistent with the new size and boundary conditions \cite{mossel10} (or lack thereof \cite{tracy08}) of the system. 
Thus, in principle,  one needs to compute the overlaps between the initial state of the system and a complete basis of Bethe states for the system without trap. Each of these Bethe states is in one-to-one correspondence with a rapidity distribution and the overlaps give the probability amplitudes for combining  those into the resultant $n_\kappa\equiv n_\kappa(t=0^+)$. 

\begin{figure}[!b]
\centering
\includegraphics[width=0.9\columnwidth]{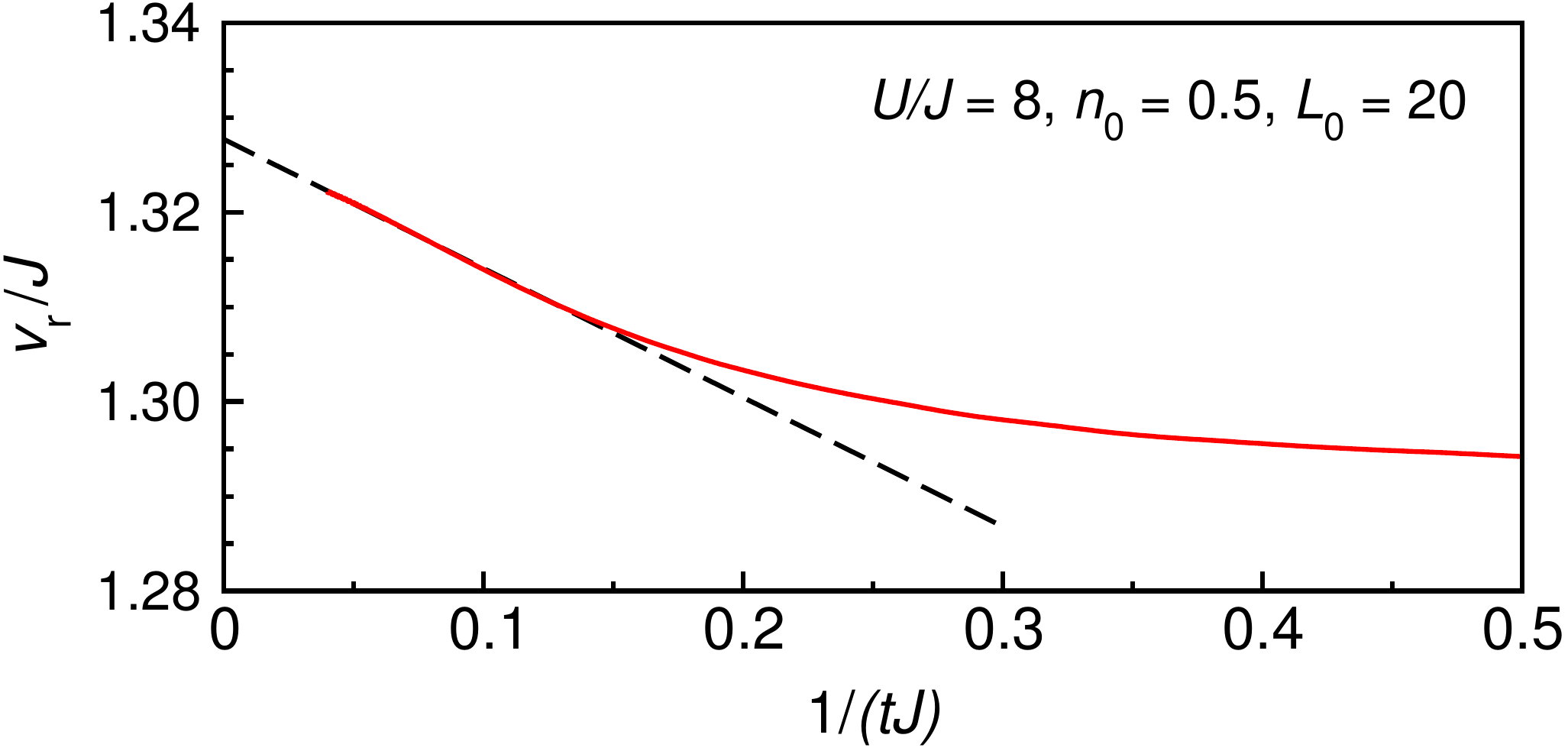}
\caption{
(Color online) Expansion velocity $v_{\rm r}$ for the Fermi-Hubbard model at $U/J=8$ with $L_0=20$ and the initial density $n_0=0.5$, as a function of the reciprocal time, obtained by tDMRG simulations (solid line).
In all tDMRG simulations, we used the time step $\Delta t J = 0.01$ and the maximal discarded weight $10^{-9}$ (different time steps were used to check convergence).
The black dashed line is a linear fit in the time range $1/(tJ) < 0.15$.
}
\label{fig:fs1}
\end{figure}

If one can identify the largest overlaps, those will give the dominant contributions, while small overlaps will give small corrections that can be left out in an approximate calculation. On the one hand, for the repulsive models we focus on in here, the initial ground states are characterized by having only real-valued charge (and, for the FHM, spin) rapidities. 
On the other hand, both real and complex rapidities (strings) are present in the full spectrum of these models, but the overlaps in the case of the latter are comparatively smaller and can be ignored in a first approximation. 
Complex rapidities are associated with different types of bound states and will most often expand more slowly, so by neglecting their contribution we  overestimate the expansion velocity of the cloud \cite{boschi14}. 
For an initial density approaching the value of one particle per lattice site, the system gets closer to a Mott-insulating state for which double occupancy is relatively suppressed (though still non-zero) and thus the projection onto bound states is also expected to be relatively suppressed.
The approximation of ignoring the bound-state contributions should thus become better the closer the trapped system is to $n_0=1$; consistent with our numerical comparisons to be shown below. To further simplify the calculation, we consider the discrete set of initial rapidities treated individually as one-particle distributions and then combine the resulting post-quench distributions to get the final result (this was already seen to correctly capture the leading contributions and give good results in other situations \cite{bolech12,campbell15}). Finally, different moments of the distribution $n_\kappa$ are combined according to Eq.~(\ref{eq:vr}) to yield the asymptotic $v_{\rm r}$.

\begin{figure}[!tb]
\centering
%   \vspace{-0.15\textheight}
\includegraphics[width=0.99\columnwidth]{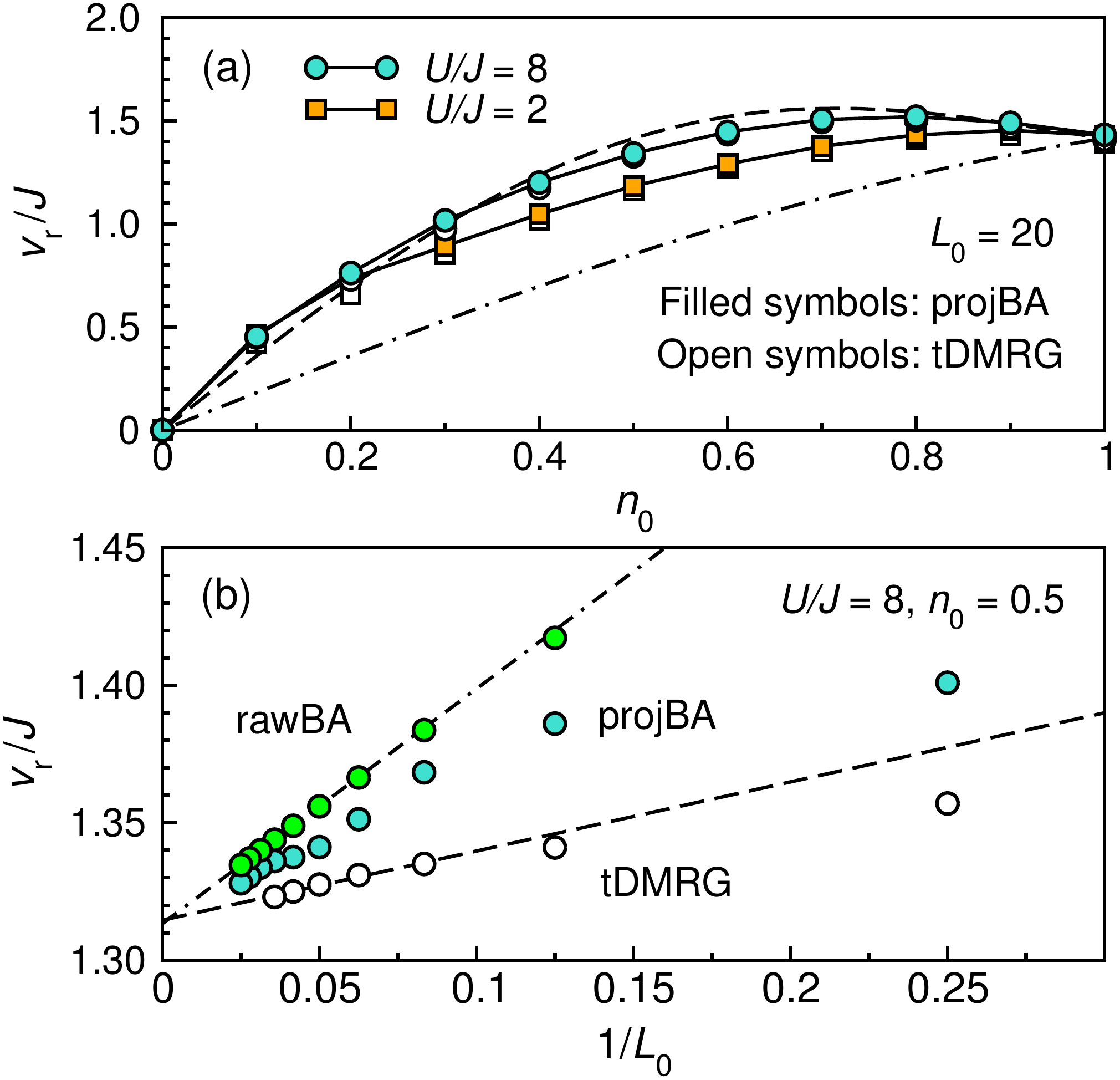}
\caption{(Color online)
{\it Expansion velocity $v_{\rm r}$ for the FHM.}
Comparison between BA (filled symbols) and tDMRG (open symbols).
The BA results are obtained with (projBA) or without (rawBA) the rapidity projection techniques.
(a) $v_{\rm r}$ versus initial density $n_0$ [$L_0=20$; $U/J=2,8$].
The dot-dashed and the dashed lines are the exact expressions for $U=0$ and $U/J=\infty$, respectively [see Eq.~(\ref{vr_n_limits})].
(b) $v_{\rm r}$ versus the initial box size $L_0$ [$U/J=8$; $n_0=0.5$].
Lines are fits to the rawBA and the tDMRG data in the range $1/L_0<0.1$.
}
\label{fig:fhm}
\end{figure}

{\it Comparison with tDMRG.}
The numerical calculation using tDMRG proceeds in a very different way. 
It performs a unitary  time evolution of the many-body wavefunction (approximated via matrix-product states~\cite{schollwoeck11}), obviously without explicitly connecting to any nontrivial integrals of motion. 
The expansion velocity of the atom cloud converges rapidly to its asymptotic form such that only relatively short times need to be reached in the simulations~\cite{langer12,vidmar13}.
From the  tDMRG simulations, one can extract the   time-dependent expansion velocity by calculating $v_{\rm r}(t) = [R(t+dt/2) - R(t-dt/2)]/dt$.
For the systems considered here, $v_{\rm r}(t)$ does not change considerably in time since the largest difference may be of the order of a few percent.
A typical time evolution of $v_{\rm r}(t)$ is shown in Fig.~\ref{fig:fs1} for the Fermi-Hubbard model  at $U/J=8$, $n_0=0.5$ and $L_0=20$.
After a few tunneling times $\propto 1/J$, we observe that $v_{\rm r}(t)$ approaches its asymptotic value as $\sim 1/(tJ)$.
We then obtain the asymptotic value of $v_{\rm r}$  by applying a linear fit
$v_{\rm r}(t) = v_{\rm r} + a/(tJ)$ in the time interval $1/(tJ) \ll 1$.

Figure~\ref{fig:fhm}(a) shows the results for the FHM for two different values of the interaction $U/J=2,8$ and $L_0=20$. 
The agreement between the tDMRG and BA calculations is generally very good.
Both approaches consistently show that the maximum values of $v_{\rm r}/J$ for strong interactions are found at initial in-trap densities $0.5<n_0<1$ 
(at $n_0=1$ one has $v_{\rm r}/J=\sqrt{2}$ regardless of the interaction strength~\cite{langer12}).
For guidance, we also show the exact results for $v_{\rm r}$ for the noninteracting and the $U/J\to\infty$ limits,  computed by taking $L_0\to\infty$~\cite{langer12},
\begin{equation} \label{vr_n_limits}
v_{\rm r}/J = \sqrt{2 \left[ 1 - \frac{\sin{(k_{\rm F})}\cos{(k_{\rm F})}} {k_{\rm F}} \right]}.
\end{equation}
Here, the Fermi momentum $k_{\rm F}$ is set by the initial density $n_0$.
For a noninteracting two-component gas,  $k_{\rm F} = n_0 \pi/2$, while for the single-component gas that describes the charge dynamics in the limit of infinite 
onsite repulsion, $k_{\rm F} = n_0 \pi$.

\begin{figure}[!tb]
\centering
%   \vspace{-0.15\textheight}
\includegraphics[width=0.99\columnwidth]{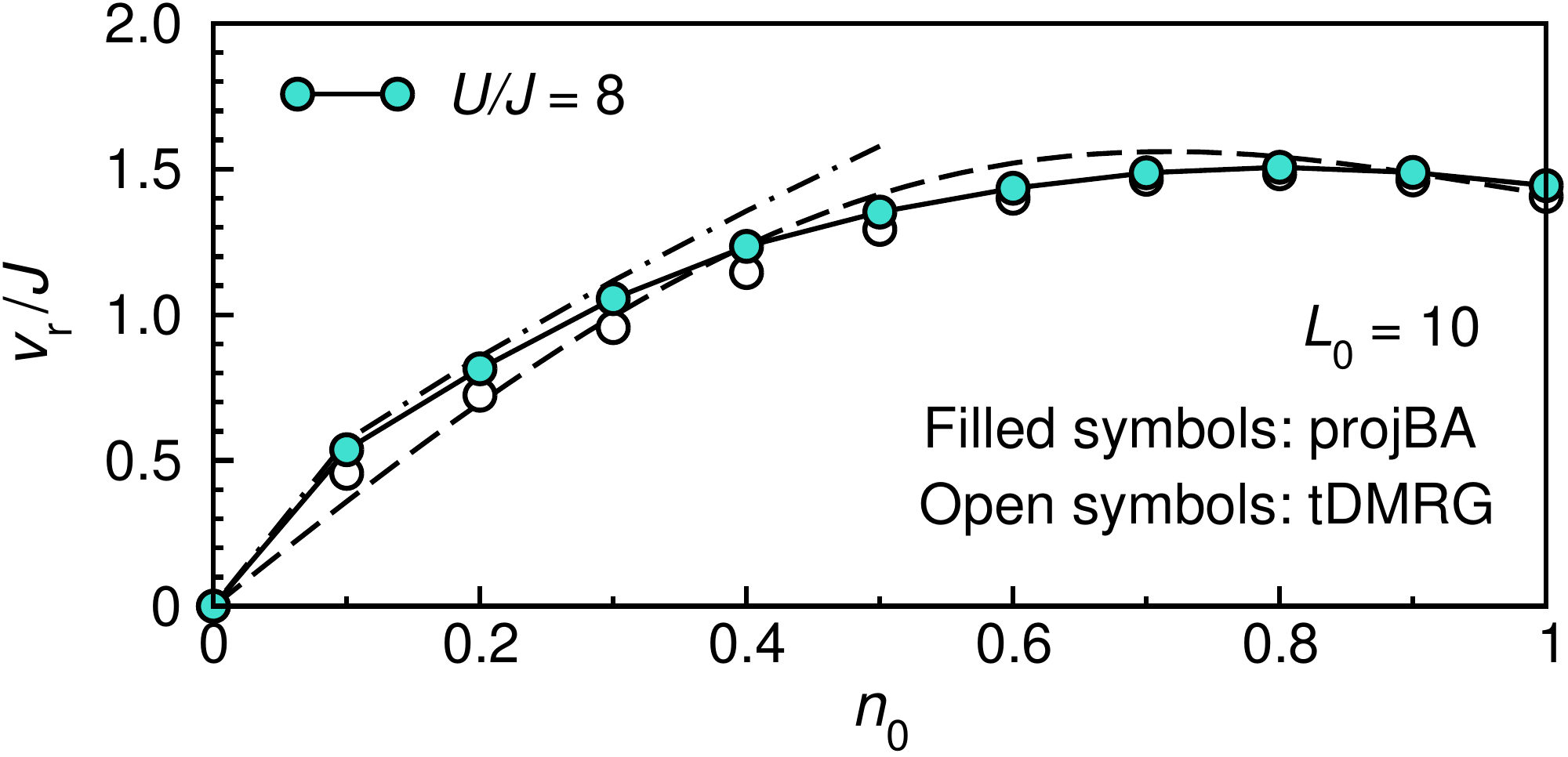}
\caption{(Color online)
{\it Expansion velocity $v_{\rm r}$ for the BHM.}
Comparison of data for $v_r$ versus $n_0$ between BA (filled symbols) using rapidity projection (projBA), and tDMRG (open symbols) [$U/J=8$, $L_0=10$]. 
The dashed line is the exact expression for $U/J=\infty$, see Eq.~(\ref{vr_n_limits}).
The dot-dashed line is the result in the dilute limit $v_{\rm r}/J=2\sqrt{E/(JN)}$.
}
\label{fig:bhm} 
\end{figure}

In Fig.~\ref{fig:fhm}(b), we compare the BA results obtained with or without the rapidity projection techniques \cite{supmat}, with the tDMRG results.
Remarkably, the finite-size scaling with respect to the initial box size $L_0$ shows that all data sets approach the same value as $L_0$ increases (see also \cite{supmat} for the BHM case). 
Thus, the approximations used in the BA-based approach become increasingly unimportant.
That the asymptotic and the scaling limits of the expansion velocities coincide is one of the main nonintuitive findings of our work.

The case of the BHM is more delicate and already within the tDMRG framework, one needs to make a (controlled) approximation by truncating the size of the local Hilbert space by introducing a maximum allowed number of bosons per site $N_{\rm cut}$.
This, as expected, works better the stronger the repulsive interaction is (we set $N_{\rm cut}=5$ for $U/J=8$).
From the point of view of BA, the system is known to be nonintegrable.
Curiously, a system of BA equations exists \cite{choy82} that yield an approximate solution which gets also more and more accurate as the interaction strength increases, and eventually becomes exact in the hard-core limit \cite{krauth91}. Coincidentally, our scheme based on the BA rapidities works also the best for strong interactions.
We can thus proceed in a similar way as for the FHM and compute the expansion velocity $v_{\rm r}$ as a function of the initial in-trap density $n_0$ of the gas.
Those results are shown in Fig.~\ref{fig:bhm} for a moderately large value of $U/J=8$ showing good agreement, in particular, for larger initial densities.
In addition, the results are quantitatively comparable to those for the fermionic case in Fig.~\ref{fig:fhm}(a) (this is obvious   
for $U/J=\infty$, see the dashed line in Fig.~\ref{fig:bhm}, where the density dynamics of fermions and bosons is identical~\cite{cazalilla11}).

The approximate BA equations for the BHM are also reliable in the dilute limit, as they tend to the corresponding equations for the Lieb-Liniger model, where integrability is restored.
In this limit (see the dot-dashed line in Fig.~\ref{fig:bhm}), it follows from Eq.~(\ref{eq:vr}) and the standard BA expression for the energy of the system that $v_{\rm r}/J=2\sqrt{E/(JN)}$, where $E$ is the energy of the system (as calculated in its prequench state and measured with respect to the bottom of the tight-binding band); in precise agreement with the exact result for the Lieb-Liniger model \cite{jukic09}.
The reason for the recovery of the exact result is two-fold:
(i) there are no bound states (and thus no complex rapidities) in the continuum limit as that is a lattice effect, and the single-rapidity approximation becomes more accurate in the dilute limit. 
In addition,
(ii) $E$ is constant after the quench. These considerations  also apply to the dilute limit of the FHM, so the same relation is expected to hold for a Gaudin-Yang gas. 
In all cases studied here, because of the repulsive interactions, the asymptotic free Hamiltonian is fermionic~\cite{vidmar13};
even for bosons, in the Lieb-Liniger case, the underlying time dependence is captured by knowing that of an antisymmetric free-fermion-like wavefunction characterized by the
values of the rapidities~\cite{gaudin14,buljan08}.
%\fabianchange{In all cases studied here,  because of the repulsive interactions, the asymptotic free Hamiltonian is fermionic \cite{vidmar13}, while for the expansion in the Lied-Liniger gas with attractive interactions, a free bosonic Hamiltonian emerges \cite{iyer12}.}

{\it Discussion.}
We showed how sudden expansion experiments, already at short times, give access to information about integrals of motion that are usually  hidden in the structure of the wavefunction.
Actual experiments of this type using cold-atom setups have already been carried out for fermions~\cite{schneider12} and bosons~\cite{kinoshita04,kinoshita06,ronzheimer13,reinhard13,vidmar15,xia15} and more accurate ones for both bosons and fermions could be  within reach exploiting single-site resolution techniques~\cite{bakr09,sherson10} (see~\cite{fukuhara13,fukuhara13a,greiner2015} for work in this direction).
The fact that most experiments use a harmonic trap does not change the picture qualitatively, it leads to a different  rapidity distribution in the initial state but after
removal of the trap, the system is again integrable.
While we have substantiated the validity of the approximations in the BA calculation by  a comparison to  tDMRG for few particles, one can push the BA calculation of asymptotic quantities to much larger initial system sizes or particle numbers~\cite{supmat}.

We speculate that systems close to an integrable point are constrained at short times by the integrals of motion of that point and reach asymptotic 
expansion states that reflect them since the gas becomes increasingly more dilute as it expands~\cite{iyer12,vidmar13}.
Therefore, perturbations from integrability have no time to generate deviations, similar to but  more robustly so than within the \textit{prethermalization} scenarios realizable on the same type of systems under different experimental conditions~\cite{berges04,dunjko12,moeckel08,eckstein09,gring12}. This conclusion is corroborated by our results for the nonintegrable BHM.
The identification and study of asymptotic expansion states seems to be a very fertile ground to explore the physics of nonequilibrium systems as they constitute a very special case of 
asymptotic states of \textit{effectively noninteracting} systems that nevertheless contain  much of the information about the character and correlations on the parent equilibrium states of the actual interacting systems. 

%\begin{acknowledgements}
{\it Acknowledgments.}
We thank N. Andrei, F. Essler,  M. Rigol and U. Schneider for useful discussions.
F.H.-M. acknowledges support from DFG Research Unit FOR 801.
L.V. was supported by the Alexander von Humboldt Foundation.
Z.M. and C.J.B. acknowledge support from DARPA's OLE program 
through ARO W911NF-07-1-0464.
This work was also supported in part by National Science Foundation 
Grant No.~PHYS-1066293 and the hospitality of the Aspen Center for Physics.

\bibliographystyle{biblev1}
\bibliography{references}

%\newpage
%\phantom{a}
\newpage
%%%%%%%%%%%%%%%%%%%%%%%%%%%%%%%%%%%%%%%%
\setcounter{figure}{0}
\setcounter{equation}{0}
\setcounter{table}{0}

\renewcommand{\thetable}{S\arabic{table}}
\renewcommand{\thefigure}{S\arabic{figure}}
\renewcommand{\theequation}{S\arabic{equation}}

\renewcommand{\thesection}{S\arabic{section}}

%\newpage

\begin{center}
\Large
\hypertarget{pagesupp}{Supplemental material}
\end{center}

\label{pagesupp}

%%%%%%%%%%%%%%%%%%
\section{Bethe-ansatz equations}\label{s2}

For reference, we give here the Bethe-ansatz equations with open boundary conditions that were used to calculate the results for the comparisons of Figs.~\ref{fig:fhm} and~\ref{fig:bhm}.
For the Fermi-Hubbard model, one has two sets of equations, one for the charge rapidities, $\kappa_{j=1,\ldots,N}$, 

\begin{equation}
\prod_{n=1}^{M}\frac{\left(\sin \kappa_{j}+\varLambda_{n}+i\frac{c}{2}\right)\left(\sin \kappa_{j}-\varLambda_{n}+i\frac{c}{2}\right)}{\left(\sin \kappa_{j}+\varLambda_{n}-i\frac{c}{2}\right)\left(\sin \kappa_{j}-\varLambda_{n}-i\frac{c}{2}\right)}=e^{2i\kappa_{j}\left(L+1\right)}
\end{equation}

\noindent (where $N$ is the number of particles and $M$ is the number of down-spin fermions in the system), and another for the spin rapidities, $\Lambda_{n=1,\ldots,M}$,

\begin{multline}
\prod_{\substack{m=1\\ m\neq n}}^{M}\frac{\left(\varLambda_{n}+\varLambda_{m}+ic\right)\left(\varLambda_{n}-\varLambda_{m}+ic\right)}{\left(\varLambda_{n}+\varLambda_{m}-ic\right)\left(\varLambda_{n}-\varLambda_{m}-ic\right)}\\
=\prod_{j=1}^{N}\frac{\left(\varLambda_{n}+\sin \kappa_{j}+i\frac{c}{2}\right)\left(\varLambda_{n}-\sin \kappa_{j}+i\frac{c}{2}\right)}{\left(\varLambda_{n}+\sin \kappa_{j}-i\frac{c}{2}\right)\left(\varLambda_{n}-\sin \kappa_{j}-i\frac{c}{2}\right)}\,.
\end{multline}
The spin and charge rapidities clearly depend on each other, yet only the charge rapidities will enter into Eq.~(6) of the main text.
%\noindent
For the Bose-Hubbard model, there are only charge rapidities and thus a single set of $N$ equations, 

\begin{multline}
\prod_{\substack{m=1\\ m\neq j}}^{N}\left(\frac{\sin \kappa_{j}-\sin \kappa_{m}-ic}{\sin \kappa_{j}-\sin \kappa_{m}+ic}\cdot\frac{\sin \kappa_{j}+\sin \kappa_{m}-ic}{\sin \kappa_{j}+\sin \kappa_{m}+ic}\right)\\
=e^{-2i\kappa_{j}(L+1)}\,.
\end{multline}

\noindent In both cases the interaction parameter is $c=U/(2J)$ and the total energy is given by the standard formula $E=-2J\sum_{i=1}^{N}\cos \kappa_{i}$.
These equations can be brought into polynomial form~\cite{bolech13} and then solved using standard root-finding procedures.
That process can be done more efficiently by solving the equations sequentially and following an iterative scheme.
Note that in the thermodynamic limit, one could have also assumed the system to be in an equilibrium thermal state and determine the rapidities by solving the corresponding Thermodynamic Bethe-Ansatz equations.

%%%%%%%%%%%%%%%%%%
\section{Finite-Size Scaling}\label{s3}

\begin{figure}[tb]
\centering
\includegraphics[width=0.9\columnwidth]{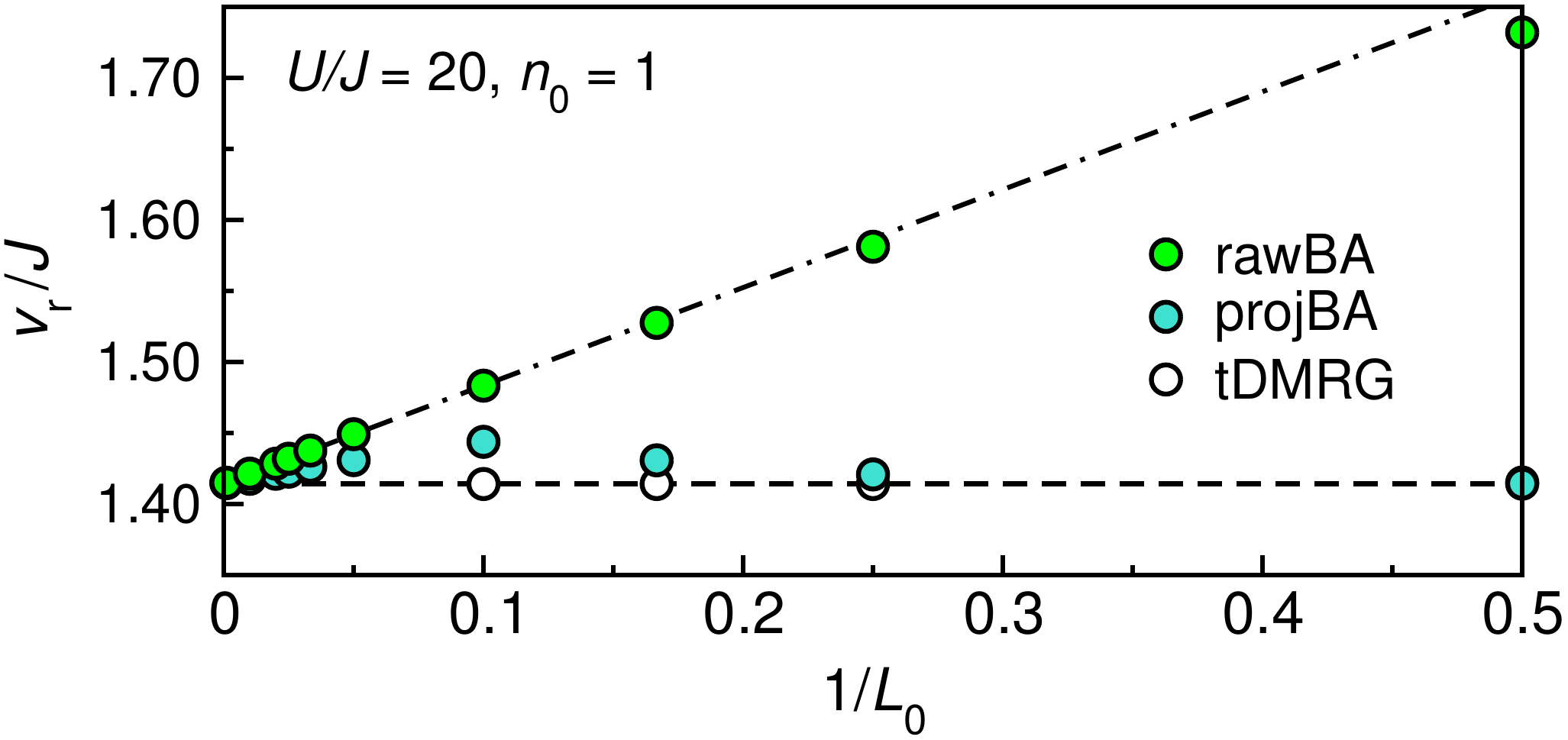}
\caption{(Color online) Expansion velocity $v_{\rm r}$ for the Bose-Hubbard model with initial density $n_0=1$ as a function of the reciprocal size of the initial system ($1/L_0$). Comparison of calculations based on Bethe ansatz without and with projection of the rapidities (filled symbols) and using tDMRG (open symbols) for a system with $U/J=20$. See Fig.~3(b) in the main text for an example of the finite-size dependence in the FHM case.
}
\label{fig:fs2}
\end{figure}

Given the size limitations of the calculations we do using tDMRG and, to a different extent, also Bethe ansatz, it is interesting to perform a finite-size scaling analysis of the results for the expansion velocities.
For concreteness, we consider the Bose-Hubbard model and focus on the case of an initially half-filled system.
As mentioned in the main text, for the $n_0=1$ case, we expect  $v_{\rm r}/J=\sqrt{2}$ from numerical studies \cite{langer12,vidmar13} and the analysis of limiting cases ($U=0$, $U/J=\infty$) \cite{langer12,vidmar13}.
In Fig.~\ref{fig:fs2} we show three different calculations, all three converging to the quoted expected result.

The three ways of calculating scale very differently with $L_0$. On the one hand, tDMRG immediately gives an accurate result for very small systems that provides up to four significant digits of precision; but for larger systems the cost of the time evolution needed to approach the ballistic expansion regime becomes larger and the error bars increase degrading the accuracy of the results. The practical size limit in this case is about $L_0\approx 20$. On the other hand, a naive Bethe-ansatz calculation gives results that, for small systems, are just order of magnitude estimates; but we understand that the reason is that one needs to project the rapidities into those of the infinite-size system. Performing this projection in the approximate way described in the main text yields up to three or four digits of precision for small systems (comparable to tDMRG), but as the system size grows, that precision degrades to only two digits, or a relative error of a few percent. This can be ascribed to the growing weight of the bound states in the expanding state of the system. Fortunately, within the Bethe-ansatz scheme we are using, we have different limitations on the size of the systems we can study and are able to consider much larger systems (in here we show up to $L_0=1000$, but even larger numbers of lattice sites are possible). The Bethe-ansatz results, with or without projection of the rapidities, scale towards the same  result. In fact, the $L_0=10$ case turns out to be the worst-case scenario in terms of accuracy and for larger systems we again achieve three or four digits of precision. A linear regression analysis in terms of $1/L_0$ yields five or more digits of precision (depending on how one restricts the data to larger system sizes).

The initial state of the Bose-Hubbard model with $n_0=1$ 
%(\fabian{Verify that this is what was meant. do we mean Bose- or Fermi- ?}) 
is also particularly interesting because we can use the properties of the Bethe-ansatz equations to derive the exact scaling expression: $v_{\rm r}/J=\sqrt{2}\sqrt{1+1/L_0}$. All the corresponding data in Fig.~\ref{fig:fs2} (green circles) agree with this expression to the level of machine precision. From there it follows that $v_{\rm r}/J\approx\sqrt{2}+(1/\sqrt{2})/L_0+\mathcal{O}(1/L_0^2)$, and we verified that not only the ordinate but also the slope of our linear regression are in good agreements.

\newpage

\end{document}